# Coupling of deterministically activated quantum emitters in hexagonal boron nitride to plasmonic surface lattice resonances

Nicholas V. Proscia[1,2,#], Robert J. Collison[3,4,#], Carlos A. Meriles[1,2,*], Vinod M. Menon[1,2,*]

[1]*Dept. of Physics, CUNY-City College of New York, New York, NY 10031, USA.*

[2]*Dept. of Physics, CUNY-Graduate Center, New York, NY 10016, USA*

[3]*Dept. of Chemistry, CUNY-City College of New York, New York, NY 10031, USA*

[4]*Ph. D. Program in Chemistry, The Graduate Center of the City University of New York, New York, NY, 10016.*

[#]*Equal contribution*

[*]Corresponding authors: cmeriles@ccny.cuny.edu, vmenon@ccny.cuny.edu

**Abstract:** Cooperative phenomena stemming from radiation-field-mediated coupling between individual quantum emitters are presently attracting broad interest for on-chip photonic quantum memories and long-range entanglement. Common to these applications is the generation of electro-magnetic modes over macroscopic distances. Much research, however, is still needed before such systems can be deployed in the form of practical devices, starting with the investigation of alternate physical platforms. Quantum emitters in two-dimensional (2D) systems provide an intriguing route because these materials can be adapted to arbitrarily shaped substrates to form hybrid systems where emitters are near-field-coupled to suitable optical modes. Here, we report a scalable coupling method allowing color center ensembles in a van der Waals material — hexagonal boron nitride — to couple to a delocalized high quality plasmonic surface lattice resonance. This type of architecture is promising for photonic applications, especially given the ability of the hexagonal boron nitride emitters to operate as single-photon sources at room temperature.

## 1 Introduction

Van der Waals materials in the form of thin sheets (1-50 nm) or single atomic layers have received a great deal of attention as a light source due to their unique mechanical, electronic and optical properties [1–4]. Particularly, their ability to conform and integrate with other material systems offer unprecedented control over their electronic and optical response [5–8]. Additionally, these material systems exhibit intriguing quantum phenomena both at the macro scale (e.g., room temperature quantum Hall effect [9], condensates [10], and superconductivity [11]) and nanoscale (e.g., single photon emission from color centers [12,13], and excitons [14–17]). These features have given the scientific community new ways to explore quantum phenomena while setting the stage for novel devices.

From among the many van der Waals systems presently attracting attention, here we focus on hexagonal boron nitride (hBN), a semiconductor featuring a wide band gap (of approximately 6 eV [18]) and high thermal conductivity [19]. This material finds a wide range of applications, from catalysis [20] to electronics (e.g., as a component for atomically-thin heterostructures [21,22]). Its crystal lattice is isomorphic to graphene, with boron and nitrogen forming the basis elements for a honeycomb structure [23]. hBN was identified recently as the host of room temperature single photon emitters (SPE) associated with point defects in its crystal lattice. These SPEs have properties needed in a single photon source (SPS) which include high excitation and out-coupling efficiency, photon purity, and a low Debye-Waller factor [69]. While the exact nature of the defects is still unclear, the low material cost, room temperature operation, high photon count rates, spectral tunability, and narrow linewidth make these emitters highly attractive for quantum technologies [24–32].

Recent studies have shown that hBN grown via chemical vapor deposition (CVD) can contain a high density of point defects that can be "activated" (i.e., transformed into a fluorescent state) either by thermal annealing or substrate-engineering [33,34]. In this latter case, the emitter position (and, to some extent, the areal concentration [35]) can be controlled, hence providing opportunities for integration with photonic structures [36,37].

Processing information with quantum emitters requires scalable methods for deterministically placing these emitters in integrated photonic circuits and subsequently coupling to structures that can direct, amplify, and potentially modulate the emitted photons. Small ensembles of independent emitters coupled to a cavity have recently gained attention with an eye on applications in quantum memory storage and its ability to act as entangled systems [38,39]. On the other hand, photonic structures such as plasmonic surface lattice resonances (SLRs) supporting delocalized or propagating modes provide one attractive route to establishing long-range interactions between electronically independent emitters so as to control emission directionality and intensity. Coupling of emitters to this class of photonic systems has attracted recent attention due to its potential as tunable light sources and compact lasing. [40–47]

In this report, we focus on engineering the emission properties of room temperature SPSs in hBN by combining deterministic emitter activation and coupling to an optically active substrate. This is achieved by placing 20-nm-thick hBN flakes on silver nano-pillar arrays which support SLRs. Capitalizing on the unique mechanical properties of 2D materials, we activate the quantum emitters located in a target area by suitably shaping the underlying substrate [7,8,33], and subsequently engineer their emission both directionally and spectrally with the aid of array-supported SLR modes [40]. While the exact mechanism by which the hBN emitters activated has not been currently isolated, a likely candidate is strain caused by the folding of the hBN around nanopillars, which in turn would lead the emitter into a bright radiative state [7,8,33,48]. The number of emitters can be controlled via the size of the activated area and the intrinsic defect density [34].

## 2 Materials and Methods

### 2.1 Sample Preparation fabrication

In our experiments, we start with the fabrication of arrays of silver nanopillars via e-beam lithography. To this end, a PMMA film (positive resist) was patterned on a glass substrate to produce regular, square arrays of cylindrical holes of variable diameter and separation (pitch). Then, a 100 nm layer of Ag was deposited on the patterned PMMA film via electron beam evaporation. Afterwards, a lift-off process was used to remove the Ag-coated PMMA via acetone, leaving behind arrays of silver nanopillars with different diameters and inter-site separations. Subsequently, a 10 nm conformal coating of alumina was deposited on the sample so as to passivate the Ag structure as well as to provide the spacing required to prevent silver-induced emitter quenching. To study the interaction between the hBN emission and the plasmonic arrays, we fabricated pillar structures with pitches varying from 220 nm to 2000 nm and with pillar diameters from 50 nm to 400 nm for each pitch. Generally, pillar arrays with a pillar-to-pillar spacing less than 100 nm were found to fuse together, hence imposing a practical limit in the type of systems we could probe. Each array was $50\times50$ $\mu m^2$ in size, large enough to allow for a collective SLR mode while still compatible with optical inspection.

The hBN used in the experiment was a CVD-grown 20-nm-thick film commercially available from Graphene Supermarket. The hBN flake had an approximate size of ~5×5mm$^2$, large enough to cover the entire patterned area with a continuous film. The hBN was deposited on the Ag pillar substrate via a wet transfer protocol described in Ref. [33].

### 2.2 Optical characterization

In order to obtain angle-resolved spectra in Figures 1 & 4, the Fourier image of the sample was projected onto the spectrometer slit (Princeton Instruments SpectraPro HR 750). The white light is focused on the array through the excitation objective (Exc. Obj.) and the transmitted light or PL was collected by the collection objective (Coll. Obj.) The collimated transmitted light is then focused by a tube lens (TL) to form a real image plane (IP) that is one focal distance away from the Bertrand lens (BL) where its back focal plane is at the spectrometer slit (Det.) [49]. In this configuration, the spectrometer captures the Fourier image of the Ag pillar array, where the linear polarizer (Pol.) is used to resolve p- and s-polarizations. For angle-resolved transmission spectra, the sample was illuminated via a tungsten lamp (Olympus) with a 0.6 NA long-working-distance objective (Olympus SLMPLN100x). For the photoluminescence spectra in Fig. 4, a beam size of ~40 $\mu m$ was used to excite the majority of the pillars in a given array.

Photoluminescence (PL) spectra were collected via an infinity-corrected 50×, 0.83 numerical aperture (NA) objective (Olympus). The confocal PL experiments in Fig. 3 were performed with a custom-built confocal microscope. A piezo nano-stage was used to raster scan the sample across a fixed laser and collection path. The focused spot size had a diameter of ~1 μm; the collection spot was ~400 nm. The PL in Fig. 3B was excited via a 510 nm, 500 fs pulsed fiber laser with a repetition rate of 80 MHz (Toptica FemtoFiber pro TVIS). The PL for the rest of the experiment was excited via a 460 nm CW diode laser.

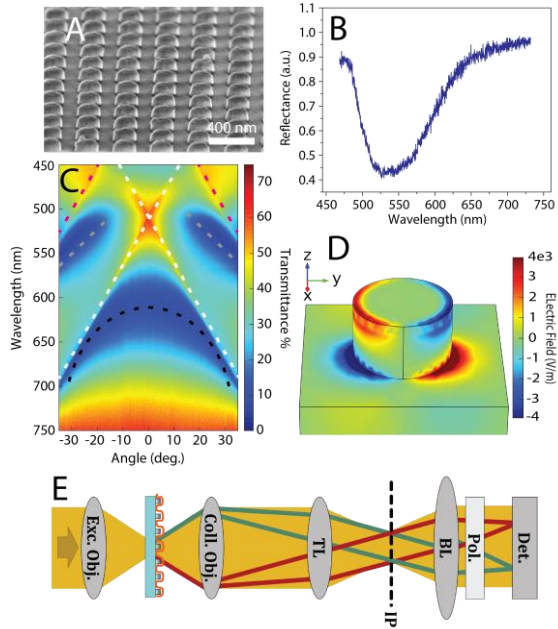

Figure 1: (A) An angled scanning electron microscopy (SEM) image of a bare array of silver nanopillars (345 nm pitch and 150 nm pillar diameter). (B) Reflectance spectrum a single pillar from a confocal collection spot. (C) Measured angle-resolved transmission spectrum of the pillar geometry in (A) for *p*-polarized transverse magnetic (TM) incident light. White and red (black and grey) dotted lines correspond to RA (SLR) modes. (D) Simulated $E_y$-field mode plot for the array in (A) of the out-of-plane SLR mode at an angle of 25 deg. and wavelength of 655 nm. (E) Schematic diagram showing the Fourier microscope used to capture the angle-resolved spectra in (C) and Fig. 4. The Bertrand lens (BL) is placed after tube lens (TL) to form an angle-resolved (signified by the green and red traces) Fourier image at the silt of the detector (spectrometer).

## 3 Results and Discussion

SLRs are hybrid plasmonic diffractive modes found in one- and two-dimensional periodic arrays of plasmonic nanoparticles. They are formed by the coupling of the localized surface plasmon resonances (LSPRs) of the discrete metal particles to in-plane diffraction orders known as Rayleigh anomalies (RAs) [50]. Fig. 1A shows an SEM image of a typical Ag nanopillar array recorded at a 70° tilt angle and a 30° scan rotation angle. It has a pitch of 345 nm and a pillar diameter of 150 nm. As seen from the SEM image, the pillars are slightly tapered at the top. This change in diameter can lead to broadening of the LSPR mode as it depends highly on the pillar cross section [40]. Hybridized SLR modes occur at wavelengths slightly red-shifted from the wavelengths of the RAs (the high transmittance lines crossing at ~510 nm in Fig. 1C). The SLR is seen in the dispersive extinction feature from 575 nm to 700 nm. The SLR in Fig. 1C probed with a broadband white source has a Fano-resonance line shape due to interference between the scattered light from the SLR mode and the directly transmitted broadband light [51]. The relevant modes here are the so-called 'out-of-plane SLRs', corresponding to LSPRs predominantly localized near the upper and lower ends of the pillars [52,53].

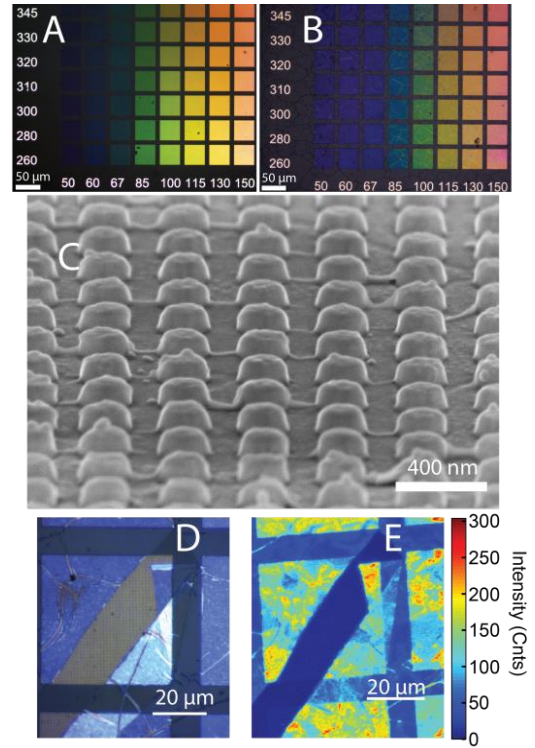

**Fig 2**. (A) Bright field optical microscopy image of various Ag pillar arrays on a glass substrate. The pitch increases vertically while pillar diameters increase horizontally. The target sizes (in nm) are visible along the left and bottom sides of the patterned region. (B) Same as in (A) but after the transfer of a 20-nm-thick sheet of hBN. (C) Angled SEM image of the Ag pillar array covered with 20 nm hBN film. The hBN film takes the shape of the substrate without tearing. (D) Optical microscope image of partially torn hBN on the Ag pillar array having 150 nm diameter pillars and a 470 nm lattice period. (E) Confocal PL image of the same array as in (D).

For the geometry in Fig. 1A, the non-dispersive LSPR mode for the 150 nm pillars, shown in Fig. 1B, splits into two branches. The branch outside the 510 nm RA has two symmetrical lobes occurring between 500 nm and 560 nm centered at the original LSPR mode (indicated by the gray dotted lines in Fig. 1C). These two lobes correspond to the SLR produced by the coupling

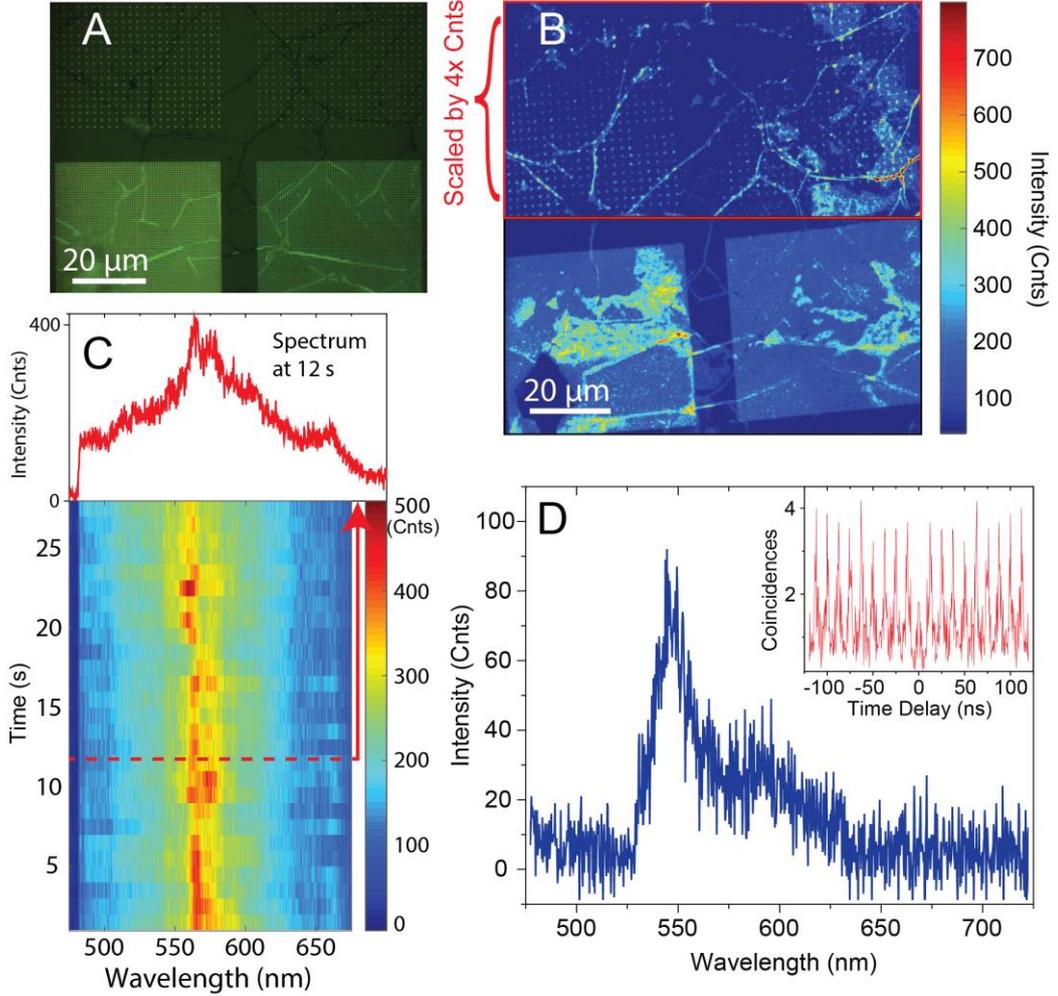

Figure 3. (A) Optical microscope image of hBN-draped Ag pillar arrays. The particle diameters are 400 and 300 nm for the pair of arrays in the left and the right halves, respectively; the lattice periods are 2 um and 620 nm for the arrays in the upper and lower halves, respectively. (B) hBN photoluminescence for the same region as in (A). For clarity, we use different color scales for the arrays in the upper and lower halves. (C) Time-resolved emission spectra from a point on the array with 430 nm pitch and 100 nm pillar diameter. Several emitters are seen alternating between bright and dark states in the 550-675 nm spectral band. The upper spectrum (red trace) corresponds to a line cut at 12 s. (D) Spectra from an isolated 100-nm-diameter pillar from a 2000-nm-pitch array. The phonon sideband is clearly visible ~165 eV from the zero-phonon line at 550 nm (blue trace). Inset: Photon correlation data via a pulsed Hanbury-Brown-Twiss measurement; an autocorrelation value $g^{(2)}(0) = 0.54$ was determined from the curve, implying 2 or less emitters.

of the Ag pillar LPSR to the higher energy $(0, \pm 1)_{air}$ RAs (red dotted lines). The lower energy $(0, \pm 1)_{glass}$ RA modes (white dotted lines) couple to the LSPR such that the resulting SLR (black dotted line) is continuous across normal incidence and follows the dispersion of the RA angles with gradually decreasing linewidths at larger angles. This is the typical behavior of SLRs [52]. The quality of the mode increases at larger angles as these modes are longer lived when they become less localized at the individual Ag pillars

The bright field microscope images of the Ag pillars arrays in Fig. 2 allow us to compare the SLR's optical properties before and after hBN transfer. Beside a ~20 nm red shift of the modes (expected from the higher index of the hBN film) and a slight decrease in the resonance quality, we find the SLRs were largely unaffected by the hBN film and retained all its core features. We fabricated Ag arrays with variable pitch and pillar diameter as seen in Figs. 2A and 2B. This allows us to study a large variety of SLR resonances as the hBN color centers occur across most of the visible spectrum, centered at ~570 nm [33,54]. The 345-nm-pitch, 150-nm-pillar-diameter array was found to have the highest emission contrast on and off the SLR. Fig. 2E shows a confocal scan of the observed PL. The hBN shows preferential emission when on the pillar array regions, while in the partially torn regions (seen in Fig. 2D) the PL intensity is drastically reduced indicating that the hBN is the source of the observed PL.

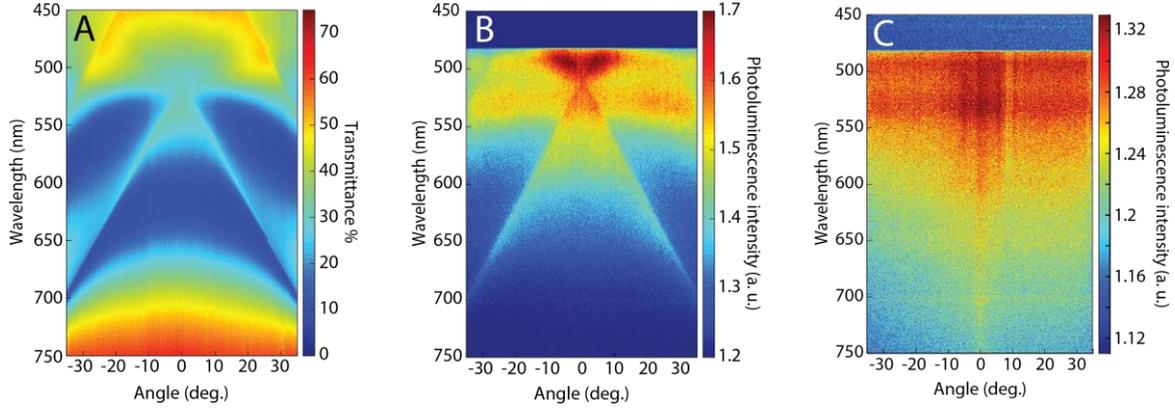

Figure 4: (A) Transverse magnetic (TM) transmission spectrum of 345-nm-period/ 150-nm-pillar-diameter array studied in Fig. 1 but with overlaid hBN. (B) Coupled PL from the hBN film on the same array with illumination from a 460 nm CW diode laser. (C) Angle-resolved PL spectrum from hBN on the 2000 nm pitch array with 150 nm diameter pillars. SLRs in the visible wavelength range are not supported due to the large spacing between the pillars.

Fig. 3 displays the emission characteristics of these hybrid hBN/Ag arrays. The hBN-induced photoluminescence from isolated plasmonic particles is highly localized to individual pillars as seen in the upper quadrants of Figs. 3A and 3B. These arrays have a 2 $\mu$m spacing and do not support collective SLRs. The pillar locations of all four arrays are clearly identifiable in the bright field image of Fig. 3A. At each pillar site, we can probe individual (or few) emitters as the pillar diameter is made sufficiently small [33]. For illustration purposes, Fig. 3D shows a spectrum for a 100-nm-diameter pillar with 2 or less defects ($g^{(2)}(0)<.54\pm.04$). The $g^{(2)}(0)$ was determined from the autocorrelation data in the inset of Fig. 3D by taking the ratio between the area under the coincidence peak at $t=0$ and the average area of 10 $t\neq0$ peaks. The area was determined by fitting each peak with a Lorentzian function [30]. Arrays that support SLRs (lower quadrants in Fig. 3A), on the other hand, become collectively bright as seen in the confocal image of Fig. 3B. As the pitch decreases, emitters get closer to each other and the collected fluorescence contains contributions from several defects. The latter leads to a broad dynamical emission across most of the visible spectrum (line cut in Fig. 3C) with emitters blinking intermittently on a timescale of seconds (contour plot in Fig 3C).

To study the interplay between emitter fluorescence and the SLRs, we look at the optical characteristics of the hBN-covered array already characterized in Fig. 1. Aside from the slight changes pointed out above, the optical transmission spectrum in Fig. 4A shows the modes remain largely unaltered. By contrast, blue laser excitation of the entire array (40 $\mu$m spot size) shows the hBN emission is drastically modified via coupling into the SLR modes (Fig. 4B). In particular we observe the PL becomes highly directional, largely mirroring the RA-induced SLR dispersion at high angles [40]. Also, the emission spectral band is extended by the SLR mode into the deep red, where the intensity of the defect emission is otherwise an order of magnitude weaker (red spectra of Fig. 3C). All of the above demonstrates good coupling between the defect center emission and the SLR modes. In the absence of an SLR mode (Fig. 4C), the hBN emission shows no dispersion. This contrast with Fig. 4B confirms that the dispersion of the hBN emission in Fig. 4B is derived from the coupling of the hBN emitters to the SLR mode supported by the 345 nm pitch Ag nanopillar array.

## 4 Conclusions

In summary, we presented a method for engineering room temperature quantum emission in a scalable platform where the emitters are activated at the array sites via substrate engineering. For suitable array parameters, the defect emission couples to the optically active plasmonic substrate and undergoes a modification in its directionality and spectral intensity. Coupling of emitter systems with photonic structures that support delocalized or propagating modes permits the mutual interaction between these distinct and spatially separated color centers. Such structures could facilitate long range energy transfer between individual emitters. Future work could focus on controlling the number of emitters coupling to the SLR mode via the size of the active region, through the number of intrinsic defects present in the material, or by spectral filtering [34].

**Acknowledgements:** N.V.P. and V.M.M. acknowledge support from the NSF MRSEC program (DMR-1420634) and the NSF EFRI 2-DARE program (EFMA -1542863). C.A.M acknowledge support from the National Science Foundation through grants NSF-1619896, NSF-1726573, and from Research Corporation for Science Advancement through a FRED Award. All authors acknowledge support from and access to the infrastructure provided by the NSF CREST IDEALS (NSF grant number HRD-1547830). The fabrication and SEM work was performed at the Advanced Science Research Center NanoFabrication Facility of the Graduate Center at the City University of New York.